\documentclass[
 aps, pra,
 amsmath,amssymb,
 11pt,
 final,
tightenlines,
 twoside,
 twocolumn,
 nofloats,
nofootinbib,
 superscriptaddress,
showkeys,
showkeywords,
 ]
{revtex4-2}

\usepackage[T1]{fontenc}
\usepackage[utf8x]{inputenc}
\usepackage[english]{babel}
\usepackage{graphicx}
\usepackage{dcolumn}
\usepackage{bm}
\usepackage[version=3]{mhchem}

\input{maik.rty}

\begin{document}

\selectlanguage{english}

\title{Isomer-selective dissociation dynamics of 1,2-dibromoethene ionized by femtosecond-laser radiation}

\author{\firstname{A.}~\surname{Mishra}}
 \affiliation{Department of Chemistry, University of Basel, Klingelbergstrasse 80, 4056 Basel, Switzerland.}

\author{\firstname{J.}~\surname{Kim}}
 \affiliation{Department of Chemistry, KAIST, Daejeon 34141, Republic of Korea.}

 \author{\firstname{S. K.} \surname{Kim}}
 \email{skkim1230@kaist.ac.kr}
 \affiliation{Department of Chemistry, KAIST, Daejeon 34141, Republic of Korea.}

 \author{\firstname{S.}~\surname{Willitsch}}
 \email{stefan.willitsch@unibas.ch}
 \affiliation{Department of Chemistry, University of Basel, Klingelbergstrasse 80, 4056 Basel, Switzerland.}

\begin{abstract}

{We study the isomer-specific photoionization and photofragmentation of 1,2-dibromoethene (DBE) under strong-field fs-laser irradiation in the gas phase complementing previous studies utilising ns- and ps-laser excitation. Our findings are compatible with a dissociative multiphoton-ionization mechanism producing a variety of ionic photofragments. Using both Stark deflection and chemical separation of the two isomers, pronounced isomer-specific photofragmentation dynamics could be observed for different product channels. While for \ce{Br+} formation, the isomer specificity appears to originate from different photoexcitation efficiencies, for the \ce{C2H2Br+} channel it is more likely caused by differences in the coupling to the exit channel. By contrast, the formation of the \ce{C2H2+} photofragment does not seem to exhibit a pronounced isomeric dependence under the present conditions. The present work underlines the importance of isomeric effects in photochemistry even in small polyatomics like the present system as well as their pronounced dependence on the photoexcitation conditions.}

\end{abstract}

\maketitle

\section{INTRODUCTION}

The relationship between molecular structure and chemical reactivity, the broad field of stereochemistry, is one of the central topics of chemical research. Different isomers of molecules exhibit distinct physical and chemical properties\cite{eliel94a,frauenfelder91a,robertsona01a} and small changes in molecular geometry can have a profound influence on the dynamics and even the outcome of chemical reactions \cite{ park02a, kim07a, taatjes13a, champenois21a, chang13a, kilaj20a, kilaj21a, dias23}. 

\begin{figure*}[!t]
\centering
\includegraphics[width= 1\textwidth]{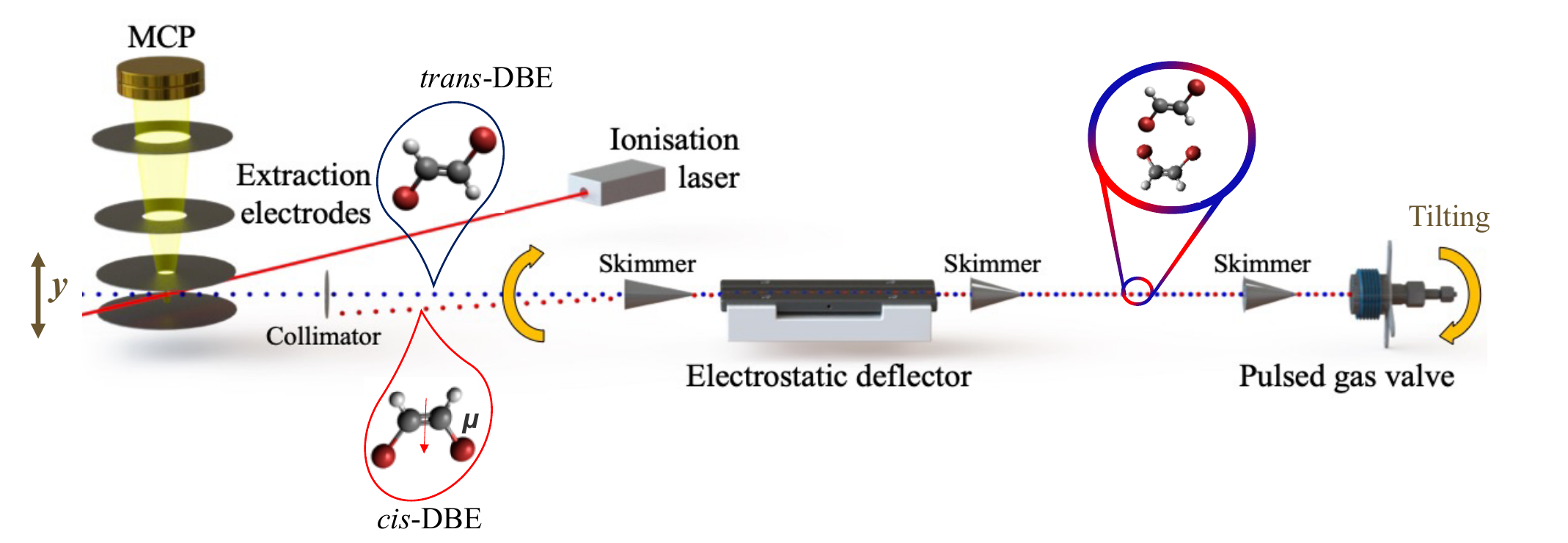} 
\caption{Schematic of the molecular-beam setup used for the electrostatic spatial separation and investigation of the  photoionization and photofragmentation of \emph{cis-} and \emph{trans-}1,2-dibromoethene (C\(_2\)H\(_2\)Br\(_2\), DBE) with a femtosecond (fs) laser at a wavelength of 775~nm.}
\label{fig1}
\end{figure*}

In this context, the photochemistry of 1,2-dibromoethene (DBE) serves as a prototypical example. DBE exhibits two geometric isomers, a polar \emph{cis} and an apolar \emph{trans} species (see insets in figure~\ref{fig1}), which classify within the C$_{2\text{v}}$ and C$_{2\text{h}}$ symmetry groups, respectively. Although the isomers of DBE can be separated by chromatography, they are not stable under ambient conditions and catalytically interconvert in the presence of atmospheric oxygen \ce{O2} \cite{steinmetz52}. 

The photolysis of bromine-substituted ethenes is of general interest because the weak C-Br bond can dissociate to form bromine atoms. For instance, in atmospheric chemistry such processes contribute to the depletion of ozone in the stratosphere\cite{garcia94}. The photochemistry of DBE has therefore been the subject of a number of studies in the past covering different wavelengths and dynamic regimes. 
The photodissociation of DBE at a wavelength of 248~nm was studied by Lee \textit{et al.} \cite{lee01} using photofragment translational spectroscopy. In their study, Br$_2^+$, Br$^+$ and C$_2$H$_2^+$ products were observed which were generated by photodissociation of the parent molecule followed by electron-impact ionization of the fragments. These products were conjectured to originate from the elimination of Br$_2$ and Br from DBE following photoexcitation with ns laser pulses. No C$_2$H$_2$Br$^+$ signals were observed in this study which was interpreted with the rapid decay of a C$_2$H$_2$Br intermediate producing Br and C$_2$H$_2$ fragments which were subsequently ionized. In a later study, Chang \textit{et al.} \cite{chang08} investigated the Br$_2$ photoproducts generated at the same photolysis wavelength using cavity-ring-down spectroscopy. Based on extensive \emph{ab-initio} calculations, they proposed different isomer-specific photodissociation pathways leading to the Br$_2$ product. The generation of the Br$_2^+$ and Br$^+$ ionic photoproducts was revisited by Hua \emph{et al.} \cite{hua11} by velocity-mapped ion imaging using ns-laser photoexcitation at 233~nm. They showed that under their experimental conditions, Br$_2^+$ was generated by dissociative multiphoton ionization of DBE rather than ionization of neutral Br$_2$ photofragments, while a range of possible pathways was discussed for the formation of the Br$^+$ fragment. 

Lipson and co-workers studied the spectroscopy and photodissociation dynamics of the individual isomers of DBE using resonance-enhanced multiphoton ionization (REMPI) and dissociation with ultraviolet photons in the range between 280 and 312.5~nm\cite{shi07a, shi07b, shi09}. The spectra observed by monitoring \ce{Br+} fragments were found to be identical for both isomers. By contrast, spectra obtained by monitoring \ce{C2H2+} fragments were found to depend markedly on the isomer. These differences were explained in terms of different intermediates arising in the relevant photofragmentation pathways leading to these products which compete in the ns-laser photoexcitation dynamics. In contrast to the preceding studies, \ce{C2H2Br+} but no \ce{Br2+} fragments were observed in these experiments.

More recently, the photofragmentation dynamics of both isomers of DBE have been studied by ps-laser multiphoton excitation with photon wavelengths in the range 232-320~nm by some of us \cite{kim21}. In contrast to the ns-laser excitation experiments covering a similar wavelength range\cite{shi09}, both \ce{C2H2Br2+} and \ce{Br2+} products were observed besides \ce{C2H2+} and \ce{Br+}. The experimental findings supported by \emph{ab-initio} calculations provided evidence for the Br$^+$ and Br$_2^+$ ionic fragments to originate from photodissociation of the ionic ground state of DBE, rather than from the ionization of neutral photoproducts. Moreover, it could be shown that the \ce{Br2+} ions originate predominantly from the \emph{cis} rather than the \emph{trans} isomer. 
These differences between the ps- and ns-laser experiments were tentatively attributed to the different multiphoton excitation dynamics in both cases. 

The recent development of techniques for the spatial separation of isomers in the gas phase by electrostatic deflection \cite{filsinger09,nielsen11} has opened up new avenues for studying isomer-specific chemistry \cite{filsinger09,chang15a, roesch14a, kilaj20a, kilaj21a, kilaj23a, Ploenes21a}. In this approach, different isomers are deflected to different extents in inhomogeneous electric fields based on their different effective dipole moments. These depend on both the permanent dipole moment in the molecular frame as well as on the rotational state of the molecule. This approach is particularly suited for separating structural isomers where one of them does not have a permanent electric dipole moment, as in the present case. 

In a previous publication \cite{mishra24}, we employed this technique to study the isomer-specific chemi-ionization of \emph{cis-} and \emph{trans-}DBE. Here, we focus on the ultrafast (fs) photoionization and dissociation dynamics of DBE with the aim of complementing the previous ns- and ps-laser studies to obtain additional information on isomeric specificities in the photochemistry of this molecule.

\section{Experimental and theoretical methods}
 \label{methods}

A detailed description of the present experimental setup has been reported previously \cite{Ploenes21a}, a schematic is displayed in figure \ref{fig1}. A molecular beam of an $\approx 1:1$ isomeric mixture of DBE (Sigma-Aldrich, 98\%) seeded in helium at 100 bar was supersonically expanded into high vacuum using an Even-Lavie pulsed gas valve\cite{even00a}. The beam was skimmed and passed through the inhomogeneous electric field of an electrostatic deflector held at a potential difference of 35 kV, where the two isomers were spatially separated based on their effective dipole moments. Different regions of the molecular beam thus exhibited different relative compositions of the isomers which were ionized using a fs-laser beam with a wavelength of 775~nm. The fs laser operated at a repetition rate of 50~Hz, a pulse duration of 150~fs, and a pulse energy of $\sim$0.2~mJ.  The fs-laser beam was focused into the vacuum chamber using a lens with a focal length of 15~cm. To avoid signal saturation and excessive fragmentation of the resulting ions, the focal point of the laser beam was offset from the center of the molecular beam by adjusting the position of the lens. 

Instead of translating the fs-laser beam to access the different regions of the deflected molecular beam, the supersonic expansion was moved by tilting the entire molecular-beam apparatus as illustrated in figure~\ref{fig1} while keeping the position of the laser beam fixed. The associated tilt angle was converted into a deflection coordinate which indicates the spatial offset of the deflected from the undeflected beam at the location of the laser focus. 

In another series of experiments the DBE isomers were first chemically separated by liquid chromatography in order to compare with the results gained by electrostatic deflection. The \emph{cis} species was then individually entrained into the molecular beam to gain additional information on its isomer-specific photodynamics. The isomeric purity of the sample amounted to $\geq$ 90\% in these measurements as confirmed by NMR spectroscopy. 

The resulting ions were electrostatically accelerated towards a microchannel plate (MCP) detector through a time-of-flight tube enabling the recording of time-of-flight mass spectra (TOF-MS). Slightly different focussing conditions were used in the experiments with the isomeric mixture and with pure \emph{cis-}DBE in order to suppress saturation.

TOF-MS were recorded at different deflection coordinates. By integrating the different peaks in the TOF-MS as a function of the deflection coordinate, deflection profiles corresponding to various ion species were obtained \cite{roesch14a, kilaj20a, Ploenes21a}. The deflection profile of the parent ion, i.e., DBE$^+$, represents a measurement of the density of DBE across the deflected molecular beam. The deflection profiles of fragment ions also encode the photofragmentation dynamics leading to their generation under the specific conditions in addition to the density distribution of the parent in the molecular beam. 

The deflection profiles were simulated using Monte-Carlo trajectory calculations \cite{filsinger09, nielsen11, chang15a, kilaj20a, roesch14a, Ploenes21a}. The dipole moment of \emph{cis-}DBE ($\mu$=1.7 D) required as input for the simulations was calculated using the B3LYP density functional method and the 6-311++G(d, p) basis set implemented in Gaussian16\cite{g16}. For each rotational quantum state of an isomer, 50'000 trajectories were computed at varying initial conditions to generate a simulated deflection profile. The profiles of individual rotational states with quantum number \textit{j} were weighted according to their thermal populations at the rotational temperature of the molecular beam and summed up to produce isomer-specific deflection profiles. Finally, the profiles of both isomers were summed up for comparison with the experimental data. 

Possible differences in the photofragmentation rate of the two isomers can affect the deflection profiles obtained from fragment-ion signals. To account for such effects, a least-squares fit was performed by introducing a relative photofragmentation rate S$_{rel}$ weighing the contributions of \emph{trans-} with respect to \emph{cis-}DBE to a specific profile. Thus, $S_{rel}>(<)~1$ indicates a stronger propensity of the \emph{trans-} (\emph{cis-}) isomer for photofragmentation in the relevant channel.

\section{Results}

\begin{figure}[!h]
\centering
\includegraphics[width=0.9\columnwidth]{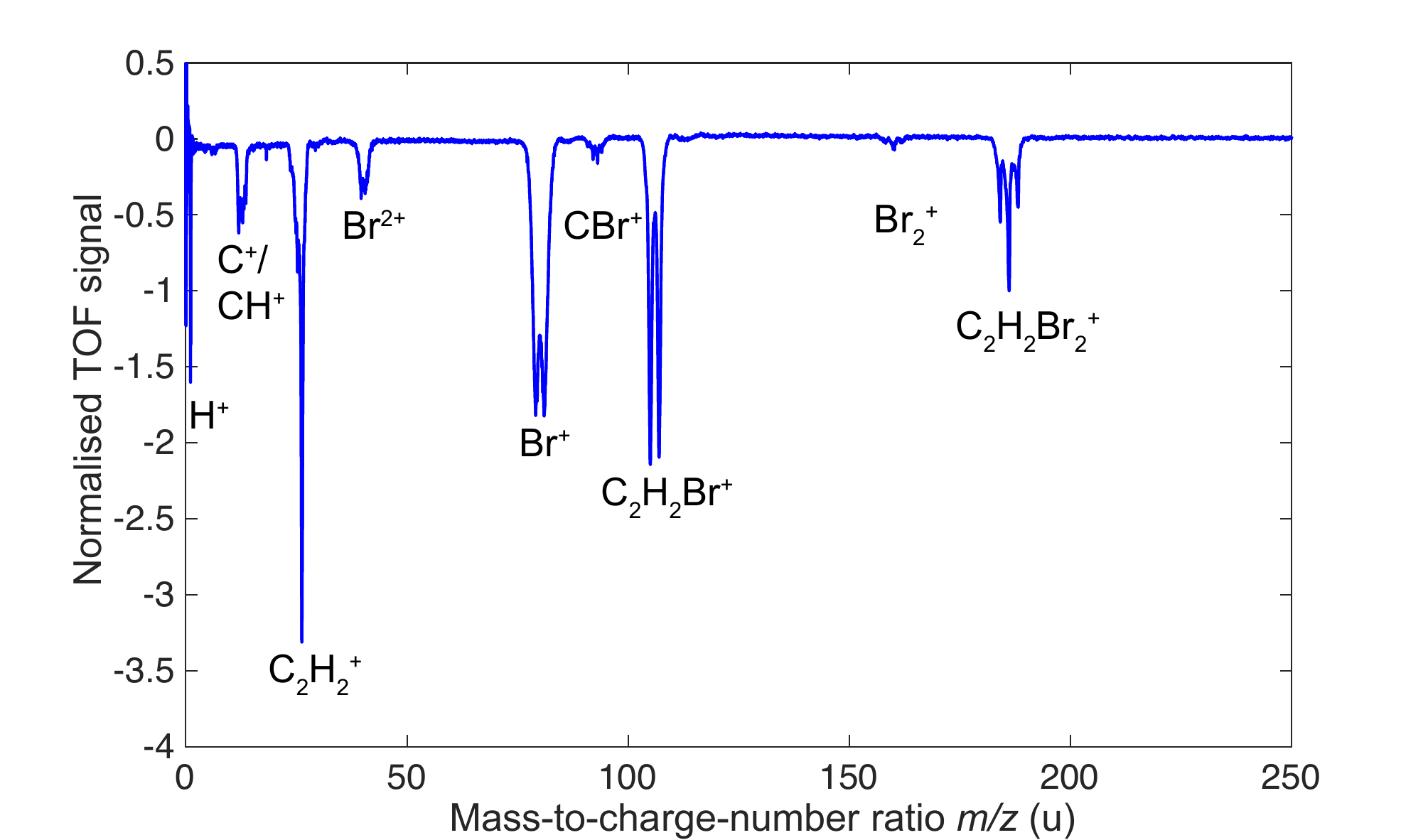} 
\caption{Time-of-flight mass spectra of DBE ionized by fs-laser radiation in the center of an undeflected molecular beam. The intensities were normalized with respect to the parent ion (\ce{C2H2Br2+}).}
\label{fig2}
\end{figure}

Figure~\ref{fig2} presents a typical TOF-MS of DBE (here the pure \emph{cis} isomer) normalized with respect to the parent ion signal (\ce{C2H2Br2+}) obtained by fs-laser ionization. The spectrum revealed four primary product-ion species around mass-to-charge ratios $m/z=$186, 106, 80, and 26~u, corresponding to \ce{C2H2Br2+}, \ce{C2H2Br+}, \ce{Br+} and \ce{C2H2+}. Furthermore, weaker signals of fragment ions were observed at $m/z=$160, 92, 40, 12, 13 and 1~u which were assigned to \ce{Br2+}, \ce{CBr+}, \ce{Br^2+}, C$^+$, CH$^+$ and H$^+$. The same signals can also be observed in the TOF-MS of the isomeric mixture.

\begin{figure*}[!t]
\centering
\includegraphics[width=0.9\textwidth]{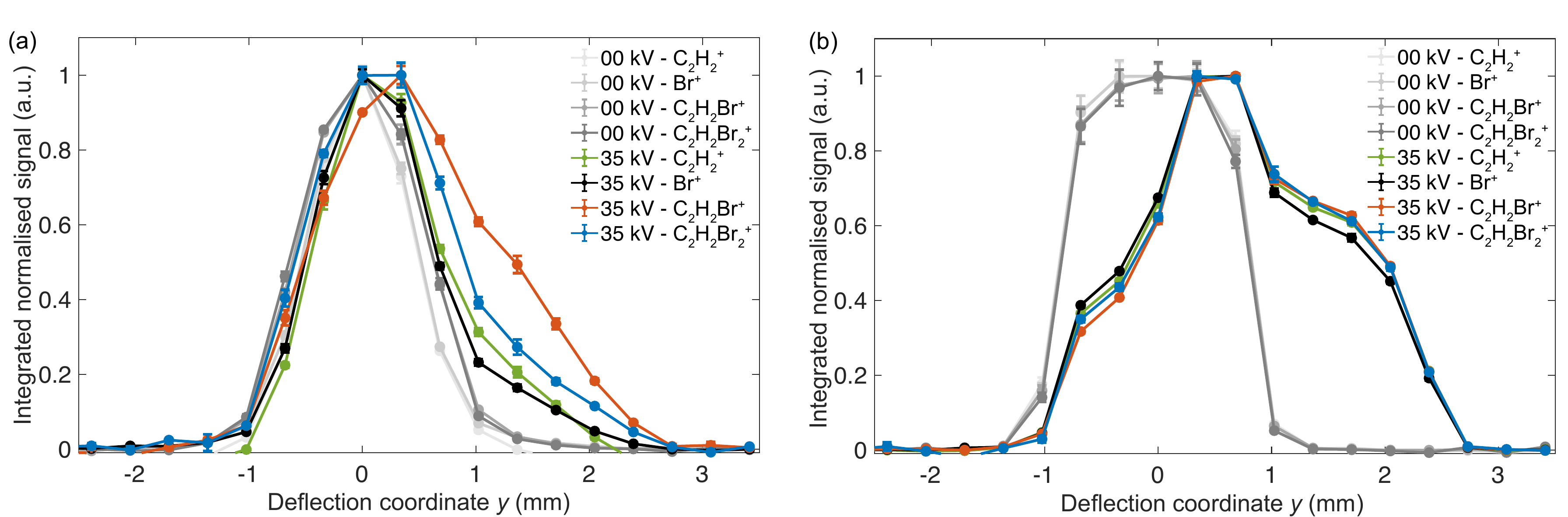} 
\caption{Deflection profiles of the parent and fragments of (a) a $\approx1:1$ mixture of the DBE isomers and (b) pure \emph{cis-}DBE obtained by fs-laser ionization. All profiles have been normalized with respect to their maxima. 
Error bars represent the standard error of three measurements with each being an average over 2000 laser shots. 
}
\label{fig3}
\end{figure*}

\begin{figure*}[!t]
\centering
\includegraphics[width=0.9\textwidth]{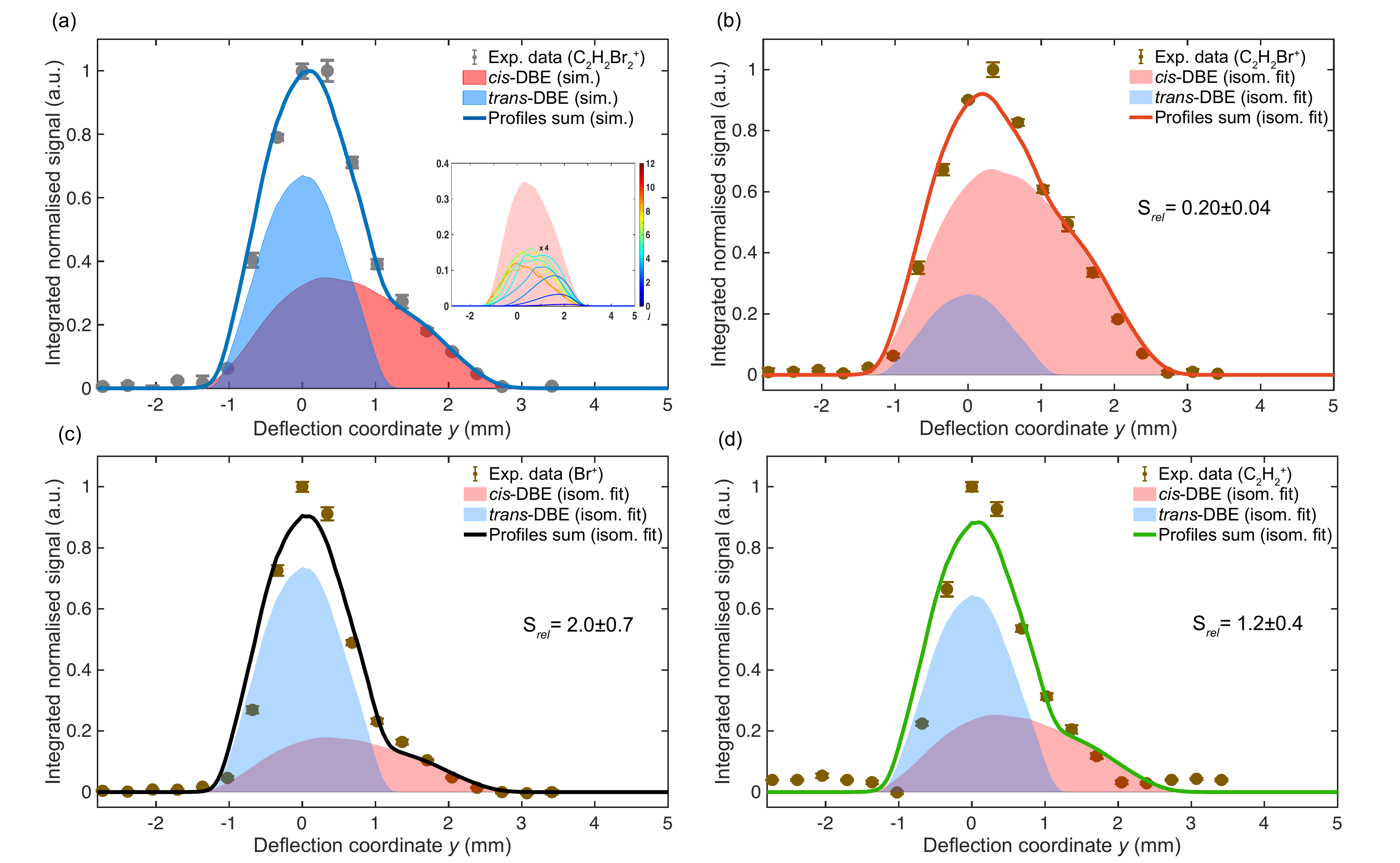} 
\caption{Deflection profiles obtained by monitoring different photofragment ions generated by fs-laser photoionization (data points) of a mixture of \emph{cis-}DBE and \emph{trans-}DBE at 35 kV deflector voltage. The blue- and red-shaded curves represent simulated deflection profiles of \emph{trans-}DBE and \emph{cis-}DBE, respectively, and the solid lines their sum. Error bars denote the standard error of three individual measurements consisting of 2000 laser shots each. The inset in (a) shows deflection profiles of individual rotational states $j$ of \emph{cis-}DBE scaled by their thermal populations and the red-shaded area represents their sum.}
\label{fig4}
\end{figure*}

\begin{figure*}[!t]
\centering
\includegraphics[width=0.9\textwidth]{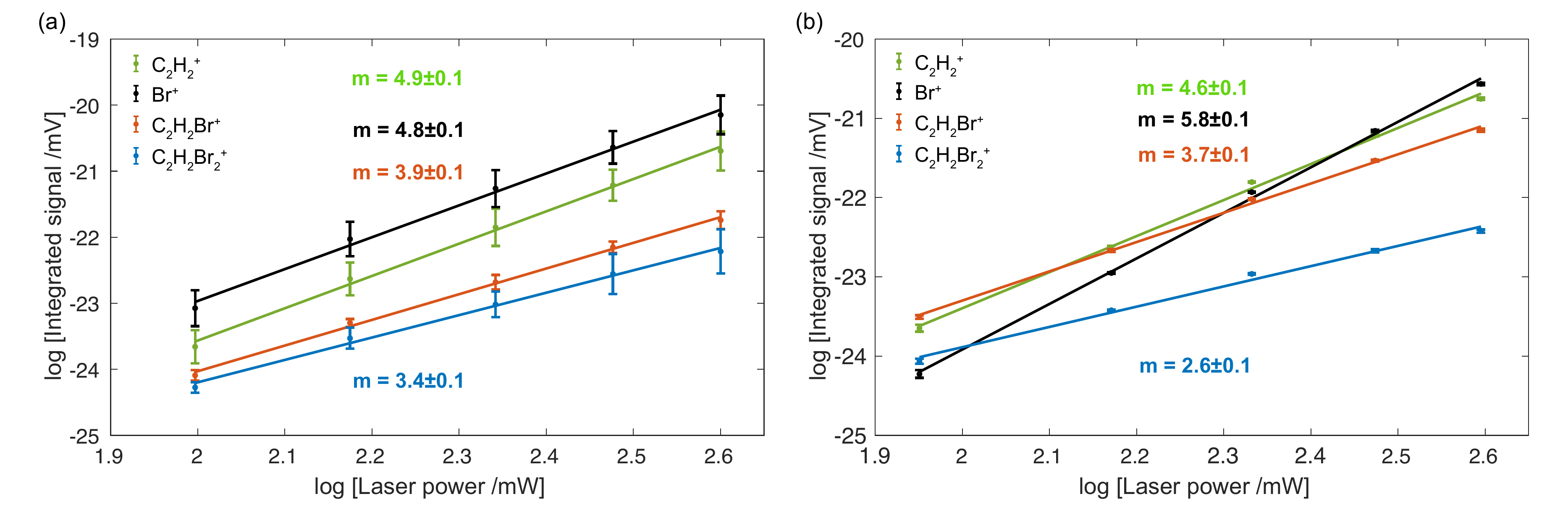} 
\caption{Doubly logarithmic representation of ion yields vs. fs-laser power for (a) an isomeric mixture of DBE and (b) pure \emph{cis-}DBE. The straight lines are linear fits to the data with $m$ denoting the slope. Uncertainties represent the standard deviations of the fits. Each data point represents the average of three measurements with each being an average over 2000 laser shots.}
\label{fig5}
\end{figure*}

Normalized deflection profiles of the dominant ion species in the TOF-MS were recorded to characterize the isomer-specific photofragmentation dynamics of \emph{cis-} and \emph{trans-}DBE under the present conditions. Figure~\ref{fig3}(a) shows the deflection profiles of ions obtained from the DBE mixture. Profiles obtained at 0~kV deflector voltage are shown in various shades of grey for the different species. Profiles obtained at 35~kV deflector voltage are plotted in blue for \ce{C2H2Br2+}, in orange for \ce{C2H2Br+}, in green for \ce{C2H2+}, and in black for \ce{Br+}. The same color code has been maintained in the subsequent figures~\ref{fig4} and \ref{fig5}.

Figures~\ref{fig3}(a) and (b) illustrate that the deflection profiles obtained without applied voltage on the electrostatic deflector exhibit a symmetric shape centered at zero deflection coordinate. In contrast, the deflection profiles obtained at a 35 kV deflector voltage are asymmetric. In figure \ref{fig3}(a), the asymmetry of the profile of the parent ion (\ce{C2H2Br2+}, blue curve) shows marked intensity at higher deflection coordinates $y\geq$ 1.5 mm suggesting that the polar \emph{cis} isomer is partially separated from apolar \emph{trans-}DBE. The extent of deflection of \emph{cis-}DBE depends on its rotational state which gives rise to different effective dipole moments in the inhomogeneous electric field of the deflector \cite{chang15a, mishra24}. Simulated rotational-state-specific deflection profiles of \emph{cis-}DBE are shown in the inset of figure \ref{fig4}(a). For clarity, curves for specific $j$ quantum numbers are displayed which were obtained by summing all corresponding profiles with different asymmetric-top pseudo-quantum number $\tau$. In general, low-$j$ states exhibit larger effective dipole moments and are, therefore, more strongly deflected thus appearing at higher deflection coordinates.

It can be seen in figure~\ref{fig3}(a) that the deflection profiles of the parent and the different fragments all differ markedly. The parent \ce{C2H2Br2+} and the fragment \ce{C2H2Br+} show a stronger deflection compared to the fragments \ce{C2H2+} and \ce{Br+}. These differences in the deflection profiles may result from different relative photofragmentation rates of either the isomers (isomeric effect) or of the different rotational states of \emph{cis-}DBE (rotational effect) - or potentially both. 

To deconvolute these effects, the corresponding deflection profiles of pure \emph{cis-}DBE at 35 kV were obtained as shown in figure \ref{fig3}(b). Interestingly, in this case all deflection profiles appear essentially identical under the present experimental conditions. This suggests that the observed differences in (a) are, therefore, not of rotational, but rather of isomeric origin.

Consequently, Monte-Carlo trajectory simulations were used to deconvolute the contributions of \emph{cis-}DBE and \emph{trans-}DBE in the deflection profiles of the various ion species, see figure~\ref{fig4}. In this figure, the data points indicate the experimental deflection profiles, while the blue and red shaded areas correspond to the simulations for \emph{trans-} and \emph{cis-}DBE, respectively, and the solid lines their sum. For the \ce{C2H2Br2+} parent (figure~\ref{fig4}(a)), the relative contribution of the curves of both isomers corresponds to their population of $\approx$1:1 in the molecular beam \cite{mishra24,shi07b}. The best match between experiment and simulation was achieved by assuming a rotational temperature $T_\text{rot}=$3.3~K in the simulation. For the fragment profiles, the relative contributions of \emph{cis} and \emph{trans} were scaled by the fit factor $S_{rel}$ capturing different photofragmentation rates in order to match the experimental data. The values of $S_{rel}$ obtained for the various fragments are indicated in figures~\ref{fig4}(b)-(d).

The results displayed in figure \ref{fig4}(b) suggest that \emph{trans-}DBE has five times reduced photofragmentation rate for the formation of the \ce{C2H2Br+} ion compared to the \emph{cis} species. By contrast, the photofragmentation rate of \emph{trans-}DBE for the formation of the \ce{Br+} product is obtained to be about twice as high as that of the \emph{cis} isomer (figure \ref{fig4}(c)). Figure~\ref{fig4}(d) shows that the formation rate of the \ce{C2H2+} product appears comparable for both isomers. The pronounced isomer-specific differences observed in the various channels suggest that different photofragmentation mechanisms are effective for the different products.

This conclusion is corroborated by measurements of the fragment-ion yield as a function of the fs-laser intensity. As the energy of a single 775 nm-photon only amounts to $\approx$1.6~eV, all ions had to be generated by multiphoton ionization and fragmentation. The ion yields exhibit a pronounced dependence on the fs-laser power, as shown in figure~\ref{fig5} (a) for the isomeric mixture and in (b) for pure \emph{cis-}DBE. The slopes $m$ obtained from a fit of the data to a linear function $y=mx+d$ in a doubly logarithmic representation indicate the apparent number of photons absorbed in the multiphoton processes leading to the different ion species. 

\begin{figure}[!t]
\centering
\includegraphics[width=\columnwidth]{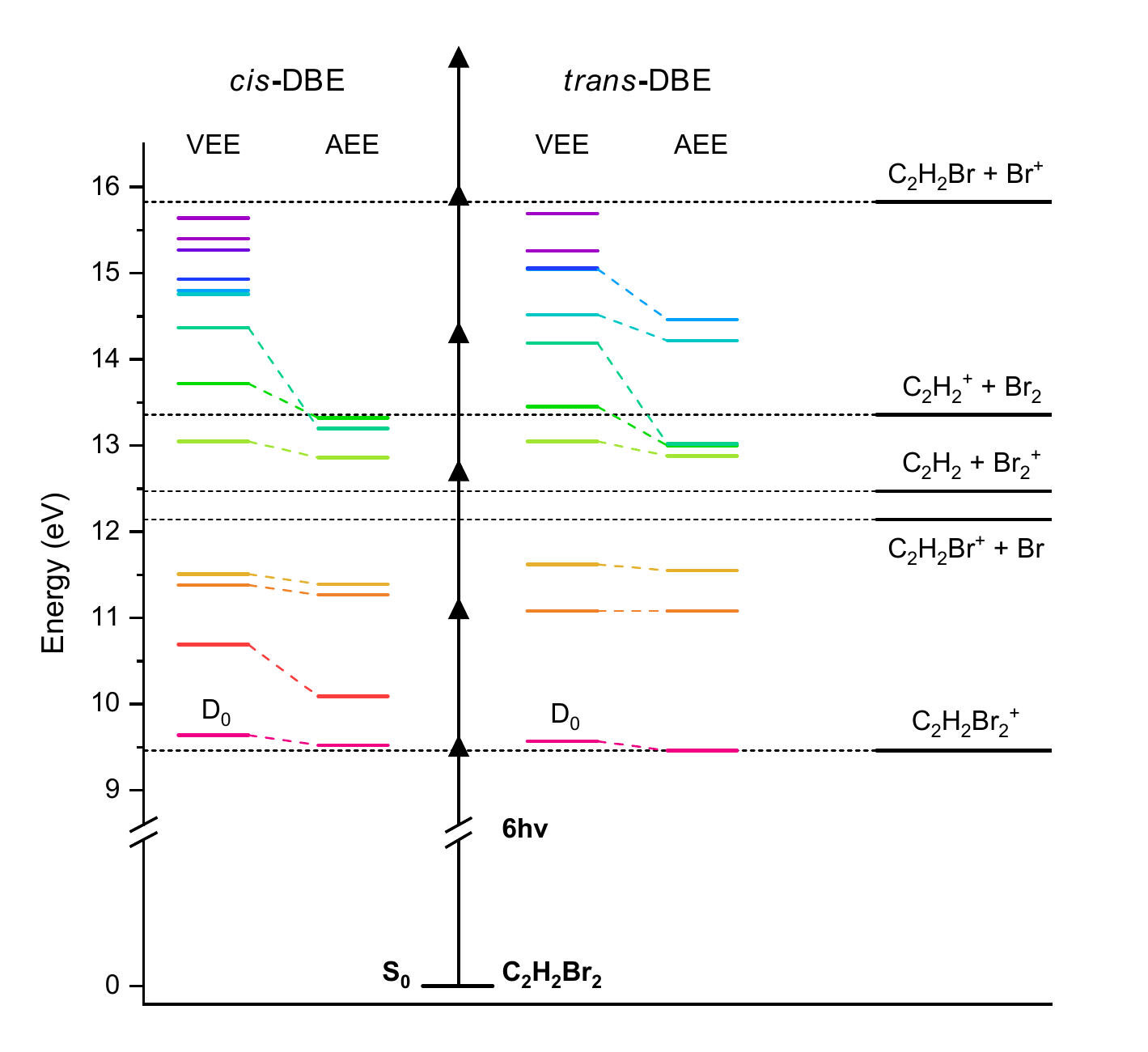} 
\caption{Cationic electronic energy levels and photofragmentation thresholds of \emph{cis}- nad \emph{trans-}DBE. The vertical (VEE) and adiabatic (AEE) excitation energies were calculated at the MP2 (ionization energy) and CAPST2 (excitation energy) levels of theory (see Ref. \cite{mishra24} for details of the calculation). Dissociation limits leading to different DI products are also indicated where the dotted horizontal line at 12.14 eV indicates the first dissociation threshold. Energies are referenced to the S$_0$ ground state of the neutral molecule. Arrows represent the energy of a fs-laser photon of 775 nm (1.6 eV).}
\label{fig6}
\end{figure}
 
The fits yielded apparent photon numbers $n=$3.4 (2.6), 3.9 (3.7), 4.9 (4.6) and 4.8 (5.8) for producing \ce{C2H2Br2+}, \ce{C2H2Br+}, \ce{C2H2+} and \ce{Br+} ions from the isomeric mixture (pure \emph{cis-}DBE). However, from energy conservation the minimum number of photons required for the formation of \ce{C2H2Br2+}, \ce{C2H2Br+}, \ce{C2H2+} and \ce{Br+} starting from neutral DBE are 6, 8, 9 and 10. This is illustrated in figure~\ref{fig6} which presents calculated vertical (VEE) and adiabatic excitation energies (AEE) of the electronic states of \emph{cis-} and \emph{trans-}DBE$^+$, along with the energies corresponding to various dissociative ionization (DI) channels (see reference~\cite{mishra24} for details of the calculations). While the apparent photon numbers follow the expected trend $n_{\text{C}_2\text{H}_2\text{Br}_2^+} < n_{\text{C}_2\text{H}_2\text{Br}^+} < n_{\text{C}_2\text{H}_2^+} < n_{\text{Br}^+}$, the discrepancy with the minimally required photon number may indicate that some intermediate absorption steps are at least partially saturated under the present experimental conditions, possibly because of (near) resonances. Note also the differences between the behaviour of the isomeric mixture in figure~\ref{fig5}(a) and pure \emph{cis-}DBE in (b): while the apparent photon numbers for producing the \ce{C2H2Br2+}, \ce{C2H2Br+} and \ce{C2H2+} are similar in both cases, there is a significant difference for the generation of the \ce{Br+} fragment (4.8 vs. 5.7 photons). This may point to the existence of additional intermediate resonances for this channel in the \emph{trans} species which are absent in the \emph{cis} isomer.

\section{Discussion}

The formation of the parent ion \ce{C2H2Br2+} can readily be explained by a multiphoton- ionization mechanism according to 
  \begin{equation}
   \ce{C2H2Br2  ->[\textit{nh}\nu] C2H2Br2^{+}} \quad n \geq 6, 
   \label{eq1}
\end{equation}
as illustrated in figure~\ref{fig6}. Because the corresponding deflection profile in figure~\ref{fig4}(a) can be simulated by maintaining the original isomeric ratio of the supersonic expansion (i.e., assuming that $S_{rel}=1$), it can be conjectured that under the present conditions the multiphoton ionization of DBE does not exhibit a pronounced dependence on the isomer.

For the formation of ionic fragments of DBE, two types of mechanisms have been discussed in the literature: dissociative ionization, where the DBE cation is first generated by multiphoton ionization and then fragmented by the absorption of additional photons~\cite{hua11, kim21}, or the parent molecule dissociates after excitation to an intermediate electronic state of the neutral and the fragments are subsequently ionized by the laser field \cite{lee01,shi09}. Given the present experimental conditions of fs-laser excitation, the latter pathway appears unlikely considering typical timescales of C--Br bond breaking (several hundreds of fs~\cite{chatterley16,toulson19}) relative to the pulse duration of the present laser (150~fs). This conclusion is in line with the previous ps-laser study \cite{kim21} which evidenced that dissociative multiphoton ionization is the dominant mechanism under conditions of short-pulse laser irradiation.

For the generation of the \ce{C2H2Br+} fragment, the value of S$_{rel}$ is 0.20 $\pm$ 0.04 extracted from figure \ref{fig4}(b) suggests that the photofragmentation rate of \emph{cis-}DBE is approximately five times larger than that of \emph{trans-}DBE. Assuming a dissociative ionization mechanism, at least two additional 775~nm photons are required to reach the \ce{C2H2Br+}+Br dissociation threshold from the ionic ground state $D_0$ (see figure~\ref{fig6}). The oscillator strengths for excitations from $D_0$ to excited cationic states $D_n, n\leq5,$ do not differ greatly between the isomers (see the supplementary information of  Ref.~\cite{kim21}) and the apparent number of photons absorbed in this channel are identical for the isomeric mixture and pure \emph{cis-}DBE within the error bars. Therefore, it can be surmised that the isomeric differences observed in the photofragmentation rate are more likely due to isomer-specific variations of the couplings of the photoexcited states to the \ce{C2H2Br+} + Br exit channel rather than the photoexcitation dynamics themselves.

Similarly, the value S$_{rel}$ = 2.0$\pm$0.7 extracted for the \ce{Br+} formation in figure \ref{fig4}(c) suggests an appreciable isomeric specificity for this fragmentation channel, albeit a smaller one than observed for \ce{C2H2Br+} generation. Assuming again a  dissociative multiphoton ionization mechanism, the absorption of at least four additional 775~nm photons from $D_0$ are required to reach the corresponding dissociation threshold (see figure~\ref{fig6}). The smaller apparent photon numbers required for this channel in the isomeric mixture ($m=4.8+0.1$) than in pure \emph{cis-}DBE ($m=5.8\pm0.1$) (see figure~\ref{fig5}) are in line with the conclusion drawn from figure~\ref{fig4}(c) that the \emph{trans} species appears more readily photodissociated in this channel compared to \emph{cis} possibly because of additional intermediate resonances. It also suggests that isomeric differences in the multiphoton excitation from $D_0$ seem to play a role here. An alternative, sequential mechanism forming \ce{Br+} from previously produced \ce{C2H2Br+} fragments seems unlikely, given the expected time scales of the laser excitation and fragementation involved. Moreover, the isomeric differences found for the \ce{C2H2Br+} and the \ce{Br+} formation were found to be markedly different, suggesting that a \ce{C2H2Br+} intermediate plays no major role for the Br$^+$ channel. 

The value S$_{rel}$ = 1.2 $\pm$ 0.4 determined from figure \ref{fig4}(d) for the \ce{C2H2+} channel indicates that the corresponding photofragmentation rates are comparable for both isomers. This is in line with similar apparent photon numbers extracted for this channel in the isomeric mixture and the pure \emph{cis} species in figure~\ref{fig5}. 

Finally, it is worth noting that the present findings of photofragmentation dynamics that is independent of the isomer for \ce{C2H2+} and dependent on the isomer for \ce{Br+} are in contrast to the previous ns-laser experiments \cite{shi09} which observed just the opposite. These differences once more underline the important role of photon energy and laser-pulse duration in the photochemistry of this molecule which in the end result in different fragmentation mechanisms: direct multiphoton dissociative ionization as conjectured to dominate the present fs-laser experiments vs. sequential fragmentation mechanisms as inferred for the conditions of the ns-laser study in Ref. \cite{shi09}.

\section{Conclusions}

In the present study, we investigated the isomer-specific photoionization and photofragmentation dynamics of 1,2-dibromoethene (DBE) under strong-field fs-laser irradiation. The experimental findings are compatible with a dissociative multiphoton ionization mechanism yielding a variety of ionic photofragments. Using both Stark deflection and chemical separation of the two isomers, pronounced isomer-specific photofragmentation dynamics could be observed for different product channels. While for \ce{Br+} formation, the isomer specificity seems to originate from different photoexcitation efficiencies, for the \ce{C2H2Br+} channel it is more likely caused by variations in the coupling to the exit channel. By contrast, the formation of the \ce{C2H2+} photofragment does not seem to exhibit a notable isomeric dependence. 

The present work complements previous studies of the photodissociation dynamics of the DBE isomers utilising ns- and ps-laser excitation. It underlines the complexity of the photochemistry even in small polyatomics like the present system, as well as its pronounced dependence on the molecular geometry and specific photoexcitation conditions. While a variety of experimental studies now exist on this photochemical model system, further insights into its isomer-specific photodynamics would critically require progress in theory, in particular concerning the multiphoton excitation dynamics of the cation and nonadiabatic couplings connecting the photoexcited states to the different product channels. 

\section*{Acknowledgements}

We thank Dr. A. Johnson, G. Holderried, G. Martin, and Ph. Kn\"opfel for technical assistance. This research was supported by the Swiss National Science Foundation under the Korean-Swiss Science and Technology Program (KSSTP) (grant nr. IZKSZ2\_188329) and the University of Basel. J.K. and S.K.K. thank the support of National Research Foundation of Korea (NRF) under project No. 2019K1A3A1A14064258. J.K. also acknowledges support from the KAIST Jang Young Sil Fellow Program.

\bibliographystyle{apsrev4-1}

\providecommand*{\mcitethebibliography}{\thebibliography}
\csname @ifundefined\endcsname{endmcitethebibliography}
{\let\endmcitethebibliography\endthebibliography}{}

\end{document}